\documentclass[a4paper,12pt]{article}
\usepackage{amssymb}
\usepackage{latexsym}
\usepackage[dvips]{graphicx}
\newcommand{\be}{\begin{equation}}
\newcommand{\ee}{\end{equation}}
\newcommand{\bea}{\begin{eqnarray}}
\newcommand{\nn}{\nonumber}
\newcommand{\eea}{\end{eqnarray}}
\begin{document}

\begin{titlepage}
\begin{flushright}
hep-th/0111047\\ UA/NPPS-13-2001
\end{flushright}
\begin{centering}
\vspace{.8in}
{\large {\bf Complex Paths \\and\\ Covariance of Hawking Radiation\\
in 2D Stringy Black Holes }}
\\

\vspace{.5in}
{\bf Elias C. Vagenas\footnote{hvagenas@cc.uoa.gr}} \\

\vspace{0.3in}
University of Athens\\ Physics Department\\
Nuclear and Particle Physics Section\\
Panepistimioupolis, Ilisia 157 71\\ Athens, Greece\\
\end{centering}

\vspace{1in}
%%%%%%%%%%%%%  ABSTRACT %%%%%%%%%%%%%
\begin{abstract}
\noindent Hawking radiation is computed in different coordinate
systems using
 the method of complex paths.
In this procedure the event horizon of the 2D Schwarzschild stringy
black hole is treated as a singularity for the
semiclassical action functional. After the regularization of the
event horizon's singularity the  emission/absorption probabilities
and the Hawking temperature in the different coordinate representations
 are derived.
The identical results obtained indicate the covariance of the
Hawking radiation.
\\
\\
\\
\\
\\
\\
\\
\\
\end{abstract}
%%%%%%%%%%%%%%%%%%%%%%%%%%%%%%%%%%%%%%

\end{titlepage}
\newpage

\baselineskip=18pt
%%%%%%%%%%%%%%%% Introduction %%%%%%%%%%%%%%%%
%\section{Introduction}
In 1974, S.W. Hawking observed that black holes radiate as a
consequence of quantum effects and their radiation spectrum is
thermal \cite{hawking}. Therefore, the  Hawking radiation extracts
energy from the black hole interior and the mass of the black hole
is diminished. This mass reduction is a physical phenomenon
independent of the coordinate representation and thus Hawking
radiation should also be covariant.
\par\noindent
S.W. Hawking described the radiation process using Bogoliubov
coefficients \cite{hawking}. Their derivation required the
knowledge of the wave modes of the quantum field in the
Schwarzschild gauge. Anyhow, solving the wave equation in an
arbitrary coordinate system in order to get the wave function \cite{diamandis} in
terms of simple functions is not always feasible. Therefore, the
problem of finding a method for evaluating  the spectrum of the
Hawking radiation without using the wave modes, turns up.
\par\noindent
J.B. Hartle and S.W. Hawking trying to solve this problem
introduced a semiclassical analysis \cite{hartle}. Applying the
Feynman path-integral approach they analytically continued the
propagator $K(x', x)$ into the complexified Schwarzschild space
where the propagator $K(x', x)$ is the amplitude for a particle to
propagate from a  spacetime point $x$ to a spacetime point $x'$ of
the Schwarzschild black hole: \be K(x', x)\approx
\sum_{paths}\exp\left[\frac{i}{\hbar}S(x', x)\right]
\label{propagator1} \ee where $S$ is the classical action for a
particular path connecting  $x$ and $x'$. This analytic
continuation led to the evaluation of the probabilities of
emission  and absorption of a particle with energy $E$: \be
P[\mbox{emission}]=P[\mbox{absorption}]\,e^{-\beta E}.
\label{spectrum1} \ee The different values of these probabilities
displayed the thermal spectrum of the Hawking radiation and its
temperature. Unfortunately the Kruskal extension used in this
method does not fit to all coordinate systems.
\par\noindent
K. Srinivasan and T. Padmanabhan introduced a semi-classical
treatment for evaluating the Hawking radiation \cite{srinivasan}
without the usage of wave modes or the complexification of spacetime.
Their method stemmed from the non-relativistic
semiclassical quantum mechanics \cite{landau}. The coordinate
singularity at the event horizon of the Schwarzschild spacetime
manifests itself as a singularity in the semiclassical propagator.
Applying the WKB approximation for the propagator and assuming
that the main contribution comes from the first term in the
expansion, the method of complex paths appropriately regularizes
the singularity in the action. Therefore, taking into account
particles tunnelling from the interior of the event horizon to the
exterior of the event horizon, the Hawking radiation is recovered.
Recently S. Shankaranarayanan, T. Padmanabhan and K. Srinivasan
applied the method of complex paths for the Schwarzschild
spacetime \cite{shanki}. They obtained the correct temperature
associated with the Hawking radiation in different coordinate
systems.
\par
In this paper, motivated by the work of S. Shankaranarayanan, T.
Padmanabhan and K. Srinivasan in the background of a Schwarzschild
black hole, we apply the method of complex paths in several
different coordinate systems describing the two-dimensional
``Schwarzschild'' stringy black hole geometry \cite{witten}.
%%%%%%%%%%%%%%%%%%%%%%%%%%%%%%%%%%%%%%%%%%%%%%%%%%%%%%%%%%%%%%%%%%%%%%%%%%%
\newline
\par\noindent
{\bf(i) Schwarzschild gauge}\par\noindent
The two-dimensional stringy black hole in the
``Schwarzschild" gauge is characterized by the line element:
\be
ds^2=-g(r)dt^2+g^{-1}(r)dr^2
\label{schwarzschild1}
\ee
where the function $g(r)$ is given by:
\be
g(r)=1-\frac{M}{\lambda}e^{-2\lambda r}
\ee
and $0<t<+\infty$, $r_H<r<+\infty$, with $
r_H=\frac{1}{2\lambda}ln(\frac{M}{\lambda})$ the
position of the event horizon of the black hole.
\newline
Considering now a massless scalar field $\Phi(t, r)$ satisfying
the Klein-Gordon equation $\Box \Phi=0$ we have in the
Schwarzschild gauge: \be -\frac{1}{g(r)} \frac{\partial^2
}{\partial t^2}\Phi (t,r) + \frac{\partial}{\partial r}\left [g(r)
\frac{\partial}{\partial r}\Phi(t,r)\right] =0. \label{kg1} \ee In
order to obtain the solution of equation (\ref{kg1}) we make the
ansantz: \be \Phi(t,r)=\exp{\left[\frac{i}{\hbar}S(t,r)\right]}
\label{ansantz1} \ee where $S$  is a function which following the
WKB approximation will be expanded in powers of $\hbar$ as
follows: \be S(t,r)=S_0 (t,r)+\hbar S_1 (t,r) +\hbar^2 S_2
(t,r)+
\ldots \, . \label{expansion1} \ee Substituting
(\ref{ansantz1}) in the Klein-Gordon equation (\ref{kg1}) we get:
\bea 0&=&-\,\frac{1}{g(r)}\left(\frac{i}{\hbar}\right)S\,
e^{\frac{i}{\hbar}S} \left(\frac{\partial^2 S}{\partial
t^2}\right)- \frac{1}{g(r)}\left(\frac{i}{\hbar}\right)^2 S^2
e^{\frac{i}{\hbar}S}\left(\frac{\partial S}{\partial
t}\right)^2-\frac{1}{g(r)}\left(\frac{i}{\hbar}\right)
 e^{\frac{i}{\hbar}S}\left(\frac{\partial S}{\partial t}\right)^2\nn\\&&+\,
\left(\frac{\partial g(r)}{\partial r}\right)
\left(\frac{i}{\hbar}\right)S e^{\frac{i}{\hbar}S}
\left(\frac{\partial S}{\partial r}\right)+
g(r)\left(\frac{i}{\hbar}\right) e^{\frac{i}{\hbar}S}
\left(\frac{\partial S}{\partial r}\right)^2 \nn\\&&+\,
g(r)\left(\frac{i}{\hbar}\right)^2 S^2 e^{\frac{i}{\hbar}S}
\left(\frac{\partial S}{\partial r}\right)^2 +
g(r)\left(\frac{i}{\hbar}\right) S\, e^{\frac{i}{\hbar}S}
\left(\frac{\partial^2 S}{\partial r^2}\right). \eea If we now use
the expansion of $S$ given in (\ref{expansion1}) and we drop all
terms of order $\hbar$ and greater, we get: \be
-\frac{1}{g(r)}\left(\frac{\partial S_0}{\partial t}\right)^2+
g(r)\left(\frac{\partial S_0}{\partial r}\right)^2 =0 \label{hj1}.
\ee
 This is
the well known Hamilton-Jacobi equation describing the motion of a
massless particle in the two-dimensional ``Schwarzschild" stringy
black hole geometry (\ref{schwarzschild1}). \par\noindent The
general solution of this equation reads:
 \be S_0 =E
\left(-t\pm\int \frac{dr}{g(r)}\right).
 \ee
 Applying the saddle point method, the
semiclassical propagator $K(x', x)$ in equation
(\ref{propagator1}) for the particle propagating from spacetime
point $x=(t_1 ,r_1)$ to spacetime point $x'=(t_2 ,r_2)$ is given
as: \be K(t_2 , r_2 ;t_1 ,r_1)=N\, \exp{\left[\frac{i}{\hbar}S_0
(t_2 , r_2 ;t_1 ,r_1)\right]} \ee where $S_0$ is the action
functional satisfying the Hamilton-Jacobi equation and $N$ is the
appropriate normalization constant. Therefore if we solve equation
(\ref{hj1}) we will be able to evaluate the amplitudes and the
probabilities of emission/absorption through the event horizon of
the two-dimensional stringy black hole.
\par\noindent
If we solve (\ref{hj1}), we get:
\bea
S_0 (t_2 ,r_2 ;t_1,r_1)&=& -E(t_2 -t_1) \pm E\int^{r_2}_{r_1}\frac{dr}{g(r)}\nn\\
&=& -E(t_2 -t_1) \pm E\int^{r_2}_{r_1}\frac{dr}{1-\frac{M}{\lambda}e^{-2\lambda r}}.
\label{solution1}
\eea
The sign ambiguity in equation (\ref{solution1}) corresponds to the
two different directions of motion of the massless particle with respect to the
event horizon of the two-dimensional black hole.
\par\noindent
For the outcoming massless particle, i.e. for $r_1 < r_H$  the
leading term $S_0$ of the action must fulfill the equation: \be
\frac{\partial S_0}{\partial r}> 0 \label{outcoming} \ee thus we
get: \bea
S_0 (t_2 ,r_2 ;t_1,r_1)&=& -E(t_2 -t_1) - E\int^{r_2}_{r_1}\frac{dr}{g(r)}\nn\\
&=& -E(t_2 -t_1) -
E\int^{r_2}_{r_1}\frac{dr}{1-\frac{M}{\lambda}e^{-2\lambda r}}
.\label{solution11} \eea Trying to evaluate the definite integral
we use the theorem of residues and choosing the contour to lie in
the upper complex plane, the leading term $S_0$ for the outcoming
massless particle is given as: \be S_0
[\mbox{emission}]=(\mbox{real part})+i\frac{\pi}{2\lambda}E.
\label{sol1} \ee For the ingoing particle, i.e. for $r_2 > r_H$
the leading term $S_0$ of the action must fulfill the equation:
\be \frac{\partial S_0}{\partial r}< 0 \label{ingoing} \ee thus we
get: \bea
S_0 (t_2 ,r_2 ;t_1,r_1)&=& -E(t_2 -t_1) + E\int^{r_2}_{r_1}\frac{dr}{g(r)}\nn\\
&=& -E(t_2 -t_1)
+E\int^{r_2}_{r_1}\frac{dr}{1-\frac{M}{\lambda}e^{-2\lambda r}}.
\label{solution2} \eea Trying to evaluate the definite integral
we use the theorem of residues and choosing the contour to lie in
the upper complex plane, the leading term $S_0$ for the ingoing
massless particle is given as: \be S_0
[\mbox{absorption}]=(\mbox{real part})-i\frac{\pi}{2\lambda}E.
\label{sol2} \ee It is well known that the probabilities are given
modulus square the amplitudes and for the case of emission and
absorption are given by: \bea P[\mbox{emission}]&\approx&
e^{\textstyle -\frac{2}{\hbar}ImS_0[\mbox{emission}]}
\label{prob1}\\
P[\mbox{absorption}]&\approx& e^{\textstyle -\frac{2}{\hbar}ImS_0[\mbox{absorption}]}
\label{prob2}.
\eea
Therefore the ratio between these probabilities is:
\be
P[\mbox{emission}]=P[\mbox{absorption}]\,e^{\textstyle -\frac{2\pi}{\hbar\lambda}E}.
\label{ratio1}
\ee
Comparing equation (\ref{ratio1}) with equation (\ref{spectrum1})
and setting $\hbar=1$ we obtain the expression for the temperature:
\be
T_H =\beta ^{-1}=\frac{\lambda}{2\pi}
\ee
which the correct formula of the Hawking temperature for
the two-dimensional stringy black hole \cite{callan}.
%%%%%%%%%%%%%%%%%%%%%%%%%%%%%%%%%%%%%%%%%%%%%%%%%%%%%%%%%%%%%%%%%%%%%%%%%%%
\newline
\par\noindent
{\bf(ii) Unitary gauge}\par\noindent The line element of the
two-dimensional stringy black hole  in the unitary gauge is: \be
ds^2=-tanh^2(\lambda y)dt^2+dy^2 \label{unidil} \ee where the
``unitary" variable $y$ is given in terms of r: \be
y=\frac{1}{\lambda}ln[e^{\lambda (r-r_H)} + \sqrt{e^{2\lambda
(r-r_H)}-1}] \ee and $0<y<+\infty$.
\newline
The Klein-Gordon equation for a massless scalar field $\Phi(t, y)$
in the unitary gauge reads:
\be -\frac{1}{\tanh^{2}(\lambda
y)}\frac{\partial ^2}{\partial t^2}\Phi(t, y) +
\frac{1}{\tanh(\lambda y)} \frac{\partial}{\partial
y}\left[\tanh(\lambda y)\frac{\partial}{\partial y}\Phi(t,
y)\right]=0. \label{kg2} \ee Substituting the ansantz
(\ref{ansantz1}) for the masssless scalar field $\Phi(t, y)$,
using the expansion (\ref{expansion1}) for the action $S$ and
neglecting all terms of order $\hbar$ and greater, the
Klein-Gordon equation (\ref{kg2}) will be: \be -\frac{1}{\tanh^2
(\lambda y)}\left(\frac{\partial S_0}{\partial t}\right)^2  +
\left( \frac{\partial S_0}{\partial y}\right)^2=0 \label{hj2} \ee
This is the Hamilton-Jacobi equation describing (in the unitary
gauge) the motion of a massless particle in the two-dimensional
``Schwarzschild'' stringy black hole. Solving equation (\ref{hj2})
we get: \be S_0 (t_2, y_2 ; t_1, y_1)=-E(t_2 -t_1) \pm
E\int^{y_2}_{y_1}\frac{dy}{\tanh(\lambda y)}. \label{solution3}
\ee The sign ambiguity as before refers to the two different
directions of motion of the massless particle with respect to the
event horizon of the black hole.
\par\noindent
For the outcoming massless particle, i.e. for $r_1 < r_H$ the
leading term
 $S_0$ of the action satisfying equation (\ref{outcoming}) and using
 the complex path method is given by:
\be S_0 [\mbox{emission}]=(\mbox{real part})+
i\frac{\pi}{\lambda}E \label{sol3} \ee while for the ingoing
massless particle, i.e. for $r_2 >r_H$  the leading term $S_0$ of
the action is given as: \be S_0 [\mbox{absorption}]=(\mbox{real
part})- i\frac{\pi}{\lambda}E. \label{sol4} \ee Comparing
equations (\ref{sol3}) and (\ref{sol4}) with (\ref{sol1}) and
(\ref{sol2}) respectively it is easily seen that there is an
additional contribution in the former case. The reason for this
fact is the double mapping of a part of the spacetime.
\par\noindent
The R-region \cite{novikov}, i.e. $r > r_H$, of the
two-dimensional stringy black hole spacetime in unitary gauge is
uniquely mapped to the corresponding R-region of the specific
spacetime in Schwarzschild gauge. The T-region \cite{novikov} ,
i.e. $r < r_H$, of the two-dimensional stringy black hole
spacetime in the unitary gauge is doubly mapped to the T-region in
the Schwarzschild gauge of the specific spacetime \cite{shanki}.
Hence, it is possible to find one point that is common to paths
contributing to absorption/emission. In the theory of complex
paths all paths are considered.
\par\noindent
Therefore taking into account all these and dividing
$S_0$'s by two, we obtain the ratio between the
corresponding probabilities of emission and absorption:
\be
P[\mbox{emission}]=P[\mbox{absorption}]\,e^{\textstyle - \frac{2
\pi}{\hbar \lambda}E} \label{ratio2} \ee which gives again the
correct formula for the Hawking temperature of the two-dimensional
stringy black hole.
%%%%%%%%%%%%%%%%%%%%%%%%%%%%%%%%%%%%%%%%%%%%%%%%%%%%%%%%%%%%%%%%%%%%%%%%%%
\newline
\par\noindent
{\bf(iii) Asymmetric gauge}
\par\noindent
The line element of the two-dimensional stringy black hole in the
asymmetric gauge is written as:
\be
ds^2=-\frac{X}{X+1}dt^2+\frac{dX^2}{4\lambda^2X(X+1)}
\label{asymmetric1}
\ee
where the ``asymmetric" variable is given by:
\be
X= e^{2\lambda(r-r_H)}-1
\ee
and $0<X<+\infty$.
\par\noindent
The Klein-Gordon equation for a massless scalar field $\Phi(t, X)$
in this gauge is: \be -\left(\frac{X+1}{X}\right)\frac{\partial
^2}{\partial t^2}\Phi(t, X) + 2\lambda
\left(X+1\right)\frac{\partial}{\partial X} \left[2\lambda
X\frac{\partial}{\partial X} \Phi(t, X)\right]=0. \label{kg4} \ee
Substituting the ansantz (\ref{ansantz1}) for the masssless scalar
field $\Phi(t, X)$,  using the expansion (\ref{expansion1}) for
the action $S$ and neglecting all terms of order $\hbar$ and
greater the Klein-Gordon equation (\ref{kg4}) becomes: \be
-\left(\frac{\partial S_0}{\partial t} \right)^2  + 4\lambda X^2
\left( \frac{\partial S_0}{\partial X}\right)^2=0 \label{hj3} \ee
This is the Hamilton-Jacobi equation describing (in the asymmetric
gauge) the motion of a massless particle in the two-dimensional
 ``Schwarzschild'' stringy black hole.
Solving equation (\ref{hj3}) we get: \be S_0 (t_2, X_2 ; t_1,
X_1)=-E(t_2 -t_1) \pm E\int^{X_2}_{X_1} \frac{dX}{2 \lambda X}\,.
\label{solution4} \ee The sign ambiguity as before refers to the
two different directions of motion of the massless particle with
respect to the event horizon of the black hole.
\par\noindent
For the outcoming massless particle, i.e. for $r_1 < r_H$ the
leading term
 $S_0$ of the action satisfying equation (\ref{outcoming}) and using the complex path method is given as:
\be S_0 [\mbox{emission}]=(\mbox{real part})+
i\frac{\pi}{2\lambda}E \label{sol5} \ee while for the ingoing
massless particle, i.e. for $r_2 > r_H$  the leading term $S_0$ of
the action is given as: \be S_0 [\mbox{absorption}]=(\mbox{real
part})- i\frac{\pi}{2\lambda}E. \label{sol6} \ee
 The ratio between the corresponding
probabilities of emission and absorption will be as before:
\be
P[\mbox{emission}]=P[\mbox{absorption}]\,e^{\textstyle -
\frac{2 \pi}{\hbar \lambda}E}
\label{ratio3}
\ee
giving again the correct formula for the Hawking temperature
of the two-dimensional stringy black hole.
%%%%%%%%%%%%%%%%%%%%%%%%%%%%%%%%%%%%%%%%%%%%%%%%%%%%%%%%%%%%%%%%%%%%%%%%%%%
\newline
\par\noindent
{\bf(iv) Painlev$\bf{\acute{e}}$ gauge}
\par\noindent
The line element of the  two-dimensional stringy black hole in the Painlev$\acute{e}$
 gauge is written as:
\be ds^2=-\left(1-\frac{M}{\lambda}e^{-2\lambda
r}\right)d\tau_{P}^2 + 2 \sqrt{\frac{M}{\lambda} e^{-2\lambda
r}}\,d\tau_{P} dr + dr^2 \label{painleve1} \ee where the
Painlev$\acute{e}$ coordinate $\tau_{P}$ \cite{painleve} satisfies
the equation: \be d\tau_{P} = dt - \frac{\sqrt{\frac{M}{\lambda}
e^{-2\lambda r}}}{\left(1-\frac{M}{\lambda}e^{-2\lambda r}\right)}
dr \ee and $0<t<+\infty$, $r_H<r<+\infty$.
\newline
Considering now a massless scalar field $\Phi(\tau_{P}, r)$
satisfying the Klein-Gordon equation we have in the
Painlev$\acute{e}$ gauge: \be \frac{\partial^2 \Phi}{\partial
\tau_{P}^{2}}-\sqrt{\frac{M}{\lambda}e^{-2\lambda r}}
\frac{\partial^2 \Phi}{\partial \tau_{P}\partial
r}-\frac{\partial}{\partial r}
\left[\left(1-\frac{M}{\lambda}e^{-2\lambda
r}\right)\frac{\partial \Phi}{\partial r}\right]
-\frac{\partial}{\partial r}
\left[\sqrt{\frac{M}{\lambda}e^{-2\lambda r}}\,\frac{\partial
\Phi}{\partial \tau_{P}}\right]=0. \label{kg5} \ee Repeating the
actions we have done in the previous gauges we obtain the
Hamilton-Jacobi equation: \be \left(\frac{\partial S_0}{\partial
\tau_{P}}\right)^2 - 2\sqrt{\frac{M}{\lambda}e^{-2\lambda
r}}\left(\frac{\partial S_0}{\partial \tau_{P}}\right)
\left(\frac{\partial S_0}{\partial r}\right)
-\left(1-\frac{M}{\lambda}e^{-2\lambda r}\right)
\left(\frac{\partial S_0}{\partial r}\right)^2 =0.
\label{hj5}
 \ee
 If we solve (\ref{hj5}), we obtain:
 \be S_0 ({\tau_{P}}_2 ,r_2 ;{\tau_{P}}_1,r_1)=
-E({\tau_{P}}_2 -{\tau_{P}}_1) \pm E\int^{r_2}_{r_1}
\left(\frac{1\pm \sqrt{\frac{M}{\lambda}e^{-2\lambda
r}}}{1-\frac{M}{\lambda}e^{-2\lambda r}}\right)dr.
\label{solution5}
\ee
The sign ambiguity is again attributed to the two different
directions of motion of the massless particle with respect to the
event horizon of the two-dimensional black hole.
\par\noindent
For the outcoming massless particle, i.e. for $r_1 < r_H$  the
leading term $S_0$ of the action must fulfill the equation
(\ref{outcoming}). In order to evaluate the definite integral we
use as before the theorem of residues and choosing the contour to
lie in the upper complex plane, the leading term $S_0$ for the
outcoming massless particle is given as: \be S_0
[\mbox{emission}]=(\mbox{real part})+i\frac{\pi}{\lambda}E.
\label{sol7} \ee For the ingoing particle, i.e. for $r_2 > r_H$
the leading term $S_0$ of the action must fulfill the equation
(\ref{ingoing}). In order again to evaluate the definite integral
we use the theorem of residues and choosing the contour to lie in
the upper complex plane, the leading term $S_0$ for the ingoing
massless particle is given as: \be S_0
[\mbox{absorption}]=(\mbox{real part})-i\frac{\pi}{\lambda}E.
\label{sol8} \ee Note that the $\frac{1}{2}$ factor that seems to
be missing in equations (\ref{sol7}) and (\ref{sol8}) is accounted
for by the fact that there is a double counting in the complex
paths \cite{shanki,novikov}.
\par\noindent
Therefore the ratio between the probabilities of emission
and absorption is the same as in all other cases: \be
P[\mbox{emission}]=P[\mbox{absorption}]\,e^{\textstyle
-\frac{2\pi}{\hbar\lambda}E}. \label{ratio4} \ee Comparing
equation (\ref{ratio4}) with equation (\ref{spectrum1}) and
setting $\hbar=1$ we obtain the right expression for the Hawking
temperature: \be T_H =\beta ^{-1}=\frac{\lambda}{2\pi} \ee for the
two-dimensional stringy black hole in the Painlev$\acute{e}$
gauge.
\par\noindent
For the case of the corresponding  charged two-dimensional stringy
black hole \cite{lee} similar results hold: The corresponding
ratio between the probabilities of emission and absorption through
the outer horizon of the charged black hole  is given as: \be
P[\mbox{emission}]=P[\mbox{absorption}] e^{\textstyle
-\frac{2\pi}{\lambda \mu}E} \label{charged} \ee
 where $E$ is the energy of a massless particle propagating in the charged two-dimensional
black hole spacetime and $\mu$ is a function of the charge $Q$ and the mass $M$ of the charged black hole.
Comparing equation (\ref{charged}) with (\ref{spectrum1}) we obtain the corresponding Hawking temperature:
\be
T_H =\frac{\lambda}{2\pi}\mu
\ee
which is the standard formula for the temperature of a charged two-dimensional stringy black hole.
These results  were also obtained by T. Christodoulakis
et al using the method of Bogoliubov coefficients \cite{theo}.
\par\noindent
It is obvious that the method introduced by the  K. Srinivasan and
T. Padmanabhan  is applicable to several static and non-static
coordinate systems of the two-dimensional stringy black hole
spacetimes. As was mentioned in the introduction the Hawking
radiation is a physical effect related to the decrease of the
black hole mass. Therefore it is expected that the energy (mass)
radiated outwards the event horizon, i.e. Hawking radiation should
be covariant. Using here the method of complex paths the thermal
spectrum and the temperature of the Hawking radiation are
recovered in different coordinate systems indicating the
covariance of the Hawking radiation of the two-dimensional stringy
black hole. We have deliberately omitted the conformal gauge in
which the event horizon is not at a finite distance and therefore
the method of complex paths cannot be applied, since there can be
no crossing of the horizon by particles. This shows one more that
the Hawking radiation is interlinked with the presence of black
hole horizons.
\par\noindent
A semiclassical approach for computing the Hawking radiation in
the Painlev$\acute{e}$ gauge was recently introduced by Parikh and
Wilczek \cite{parikh}. This method considers the Hawking radiation
as a pair creation outside the event horizon with the negative
energy particle tunnelling into the black hole. We have performed
the method of Parikh and Wilczek for the charged two-dimensional
stringy black hole background \cite{elias}. In the present paper
we have also taken into consideration the contribution to the
amplitudes of the pair creation from within the event horizon.
%%%%%%%%%%%%%%%%%%%%%%%%%%%%%%%%%%%%%%%%%%%%%%%%%%%%%
%%%%%%%%%%  ACKNOWLEDGEMENTS %%%%%%%%%%%%%%%%%
\section*{Acknowledgements}
The author would like to thank Ass. Professor T. Christodoulakis
for useful discussions and enlightening comments. This work is
financially supported in part by the University of Athens' Special
Account for the Research.
The author would also like to thank S. Shankaranarayanan for bringing
to the author's attention his recent work which was the original motivation
 for the present work.

%%%%%%%%%%%%%%% BIBLIOGRAPHY  %%%%%%%%%%%%%%%%%

%%%%%%%%%%%%%%%%%%%%%%%%%%%%%%%%%%%%%%%%%%%%%%%%%%%%%%%%%%%%%%%%%%%%%%%%%%%%%%


\begin{thebibliography}{99}

\bibitem{hawking} S.W. Hawking, Comm. Math. Phys. {\bf43} (1975) 199.

\bibitem{diamandis} C. Chiou-Lahanas, G.A. Diamandis, B.C. Georgalas,
 X.N. Maintas, E. Papantonopoulos, Phys. Rev. D {\bf52} (1995) 5877.

\bibitem{hartle} J.B. Hartle, S.W. Hawking, Phys. Rev. D {\bf13} (1976) 2188.

\bibitem{srinivasan} K. Srinivasan, T. Padmanabhan, Phys. Rev. D {\bf60}
(1999) 24007.

\bibitem{landau} L.D. Landau, E.M. Lifshitz, {\it Quantum Mechanics
(Non-relativistic Theory)}, Course of Theoretical Physics, Volume 3,
Pergamon Press, New York, 1975.

\bibitem{shanki} S. Shankaranarayanan, K. Srinivasan, T. Padmanabhan ,
Mod. Phys. Lett. A {\bf16} (2001) 571 ; S. Shankaranarayanan, T.
Padmanabhan , K. Srinivasan, Class. Quant. Grav. {\bf19} (2002)
2671.

\bibitem{witten} E. Witten, Phys. Rev. D{\bf44} (1991) 314;
G. Mandal, A.M. Sengupta and S.R. Wadia,
Mod. Phys. Lett. A {\bf6} (1991) 1685.

\bibitem{callan} C.G. Callan, S.B. Giddings, J.A. Harvey and
A. Strominger, Phys. Rev. D {\bf45}, (1992) R1005;
S.B. Giddings and W.M. Nelson, Phys. Rev. D {\bf46}
(1992) 2486.

\bibitem{novikov} I.D. Novikov, Communications of Shternberg
State Astronomical Institute, {\bf132} (1964) 3.

\bibitem{painleve} P. Painlev$\acute{e}$, C.R. Acad. Sci.(Paris)
{\bf173} (1921) 677.

\bibitem{lee} H.W. Lee, Y.S. Myung, J.Y. Kim and D.K. Park,
Class. Quant. Grav. {\bf14} (1997) L53.

\bibitem{theo} T. Christodoulakis, G.A. Diamandis, B.C. Georgalas,
E.C. Vagenas, Phys. Lett. B {\bf501} (2001) 269.

\bibitem{parikh} M.K. Parikh and F. Wilczek, Phys. Rev. Lett.
{\bf85} (2000) 5042.

\bibitem{elias} E.C. Vagenas, Phys. Lett. B {\bf503} (2001)
399.

\end{thebibliography}
 \end{document}